\documentclass[aps,prc,preprint,superscriptaddress]{revtex4-1}

\usepackage{color}
\usepackage{graphicx}
\usepackage{hyperref}

\newcommand{\jpsi}{\mathrm{J/}\psi}

\newcommand{\WgPb}{W_{\gamma\mathrm{Pb}}}
\newcommand{\gA}{\gamma\mathrm{A}}
\newcommand{\WgA}{W_{\gA}}
\newcommand{\Wgp}{W_{\gamma\mathrm{p}}}
\newcommand{\sNN}{\sqrt{s_{\rm NN}}}

\begin{document}



\title{\boldmath
Coherent and incoherent $\mathrm{J/}\psi$  photonuclear production in an energy-dependent hot-spot model
\unboldmath}


\author{J. Cepila}
\affiliation{Faculty of Nuclear Sciences and Physical Engineering,
Czech Technical University in Prague, Czech Republic}
\author{J. G. Contreras}
\affiliation{Faculty of Nuclear Sciences and Physical Engineering,
Czech Technical University in Prague, Czech Republic}
\author{M. Krelina}
\affiliation{Departamento de F\'{\i}sica, Universidad T\'ecnica Federico Santa Mar\'{\i}a;
Centro Cient\'{\i}fico-Tecnol\'ogico de Valpara\'{\i}so-CCTVal,
 Casilla 110-V, Valpara\'{\i}so, Chile}
\affiliation{Faculty of Nuclear Sciences and Physical Engineering,
Czech Technical University in Prague, Czech Republic}


\date{\today}

\begin{abstract}
In a previous publication, we have presented a model for the photoproduction of $\jpsi$ vector mesons off protons, where the proton structure  in the impact-parameter plane is described by an energy-dependent hot-spot profile. Here we extend this model to
 study the photonuclear production of $\jpsi$ vector mesons in coherent and incoherent interactions of heavy nuclei. We study two methods to extend the model to the nuclear case: using the standard Glauber-Gribov formalism and using geometric scaling to obtain the nuclear saturation scale. We find that the incoherent cross section changes sizably with the inclusion of subnucleonic hot spots, and that this change is energy dependent.  We propose to search for this behavior by measuring the ratio of the incoherent to coherent cross section at different energies. We compare the results of our model to results from RHIC and from the Run 1 at the LHC finding a satisfactory agreement. We also present predictions for the LHC at the new energies reached in Run 2. The predictions include $\jpsi$ production in ultra-peripheral collisions,  as well as the recently observed photonuclear production  in peripheral collisions.
\end{abstract}


\maketitle

\section{Introduction}
Diffractive photoproduction of $\jpsi$ vector mesons in high-energy interactions is a sensitive probe of the gluon distribution of hadrons in the small-$x$ region, where saturation effects are expected to be important~\cite{Albacete:2014fwa}. As such, it has been extensively studied experimentally at the LHC~\cite{Contreras:2015dqa,Andronic:2015wma} and it is an essential component of the research program envisaged for future facilities~\cite{Accardi:2012qut, AbelleiraFernandez:2012cc}.

 The idea of describing the structure of protons with subnucleonic degrees of freedoms representing regions of high-gluon density, so-called hot spots, has recently yielded very interesting results.
 In particular, it has been found that, using a Good-Walker formalism~\cite{Good:1960ba}, the dissociative photoproduction of $\jpsi$ off proton targets was  sensitive to geometric fluctuations of the positions of hot spots in the impact-parameter plane of the interaction~\cite{Mantysaari:2016ykx,Mantysaari:2016jaz}. In this model, the proton is made up of three hot spots, and the comparisons to experimental data is performed at a fixed center-of-mass energy ($\Wgp$) of the photon-proton system.  Recently, this model has also been extended to the case of photonuclear production at the LHC~\cite{Mantysaari:2017dwh}.

In a further development, a model in which the number of hot spots increased with decreasing $x$ was presented in~\cite{Cepila:2016uku}. Such a change in the transverse profile of the
target is qualitatively similar to the formal results obtained by evolving the JIMWLK
equation and depicted in Fig. 1 of~\cite{Schlichting:2014ipa}.
The model presented in~\cite{Cepila:2016uku} was able to correctly describe experimental data on the $\Wgp$ dependence of both, the exclusive and the dissociative production of $\jpsi$.
In this model, the energy dependence of the dissociative process shows
a sharp decrease, quantitatively different than expectations from extrapolations of
HERA data, providing a new signature of saturation effects, which could be measurable
at current LHC energies~\cite{Cepila:2016uku}.

Here, we extend our model from~\cite{Cepila:2016uku} to the case of photonuclear interactions.
We follow two different approaches to go from proton to nuclear targets. One uses the standard Glauber-Gribov formalism as proposed in~\cite{Armesto:2002ny}, while the other relies on geometric scaling  to compute the nuclear saturation scale using as an input the saturation scale in the proton~\cite{Armesto:2004ud}.

 In agreement with the results of~\cite{Mantysaari:2017dwh}, we show that the incoherent photonuclear production of $\jpsi$ is sensitive to the hot spot structure of nucleons. In addition, we predict that this effect is energy-dependent and that this dependence could be observed by measuring the ratio of the incoherent to the coherent photonuclear cross section as a function of the energy of the interaction.

We compare the results of our model with data from RHIC and from the Run 1 at the LHC and make predictions for measurements to be performed at the LHC using Run 2 data, including  the prediction of $\jpsi$ photonuclear production in both, peripheral and ultra-peripheral interactions.

These studies of fluctuations in the transverse structure of hadronic targets have been
shown to be relevant in other contexts besides the diffractive photonuclear production
of vector mesons. For example, similar ideas have been applied to explain the hollowness
effect in proton-proton interactions at high energies~\cite{Albacete:2016pmp}. Another area where
such models have been applied is the initial state of collisions involving relativistic
nuclei where effects of initial spatial asymmetries may impact on key measurements
and their interpretations, e.g.~\cite{Welsh:2016siu,Mantysaari:2017cni}.

\section{Description of the formalism
\label{sec:form}}

\begin{figure*}
\includegraphics[width=0.48\textwidth]{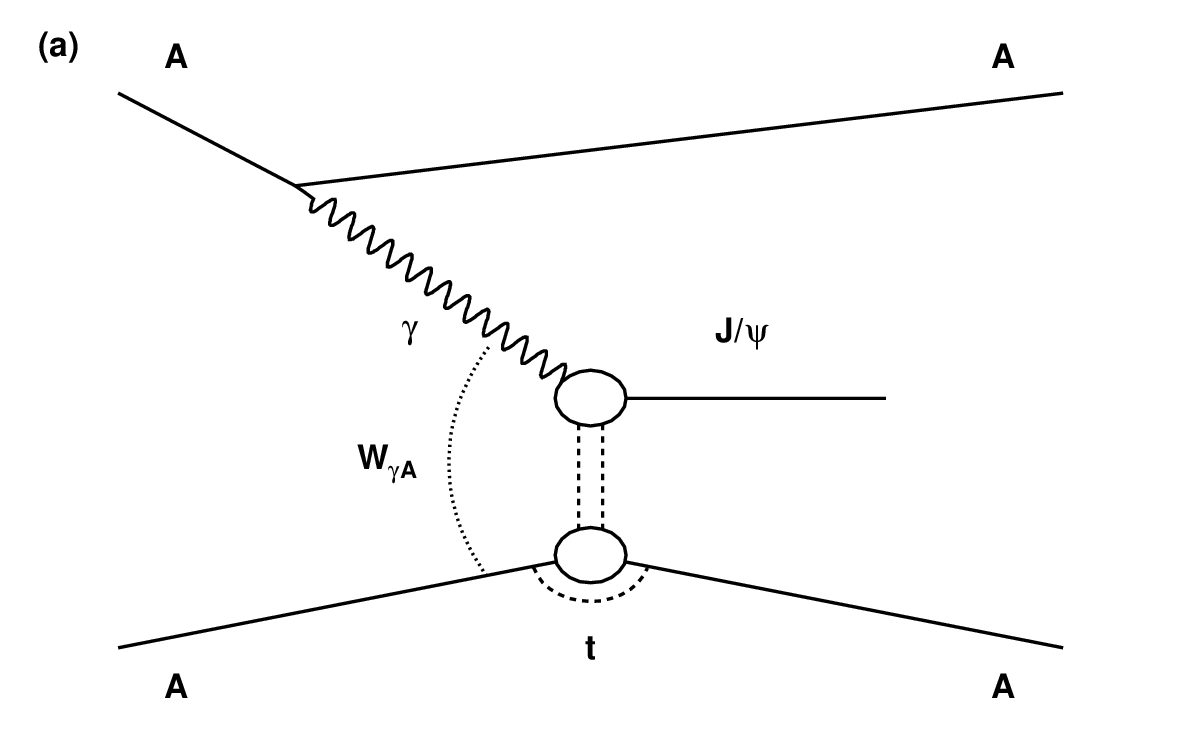}%
\includegraphics[width=0.48\textwidth]{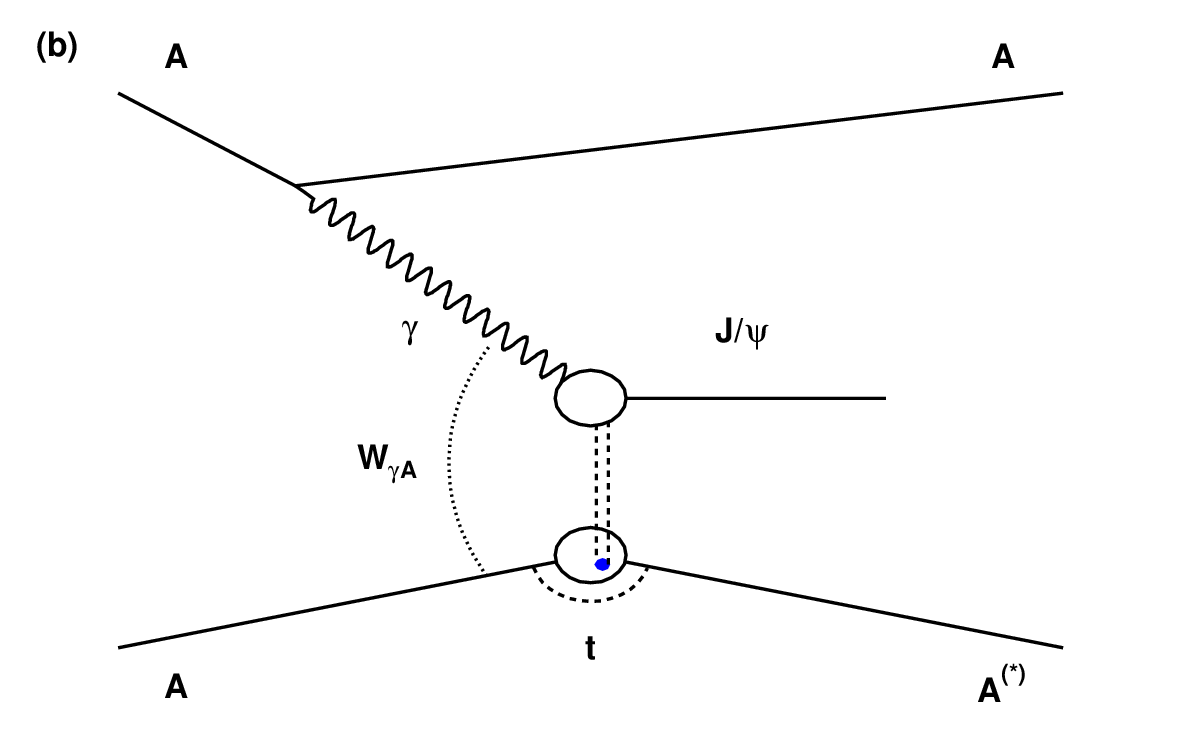}
\caption{\label{fig:diag}  Diffractive photonuclear production of $\jpsi$ in A-A collisions. In diagram (a) the photon interacts with the full nucleus A, while in (b) it scatters off one of the nucleons in A. Accordingly these processes are called coherent and incoherent $\jpsi$ production, respectively.}
\end{figure*}

The processes which we study here are shown in Fig~\ref{fig:diag}. Two heavy nuclei A approach at high energies. One of the nuclei coherently emits a quasi-real photon, which collides with the other nucleus. We are interested in  the following two processes: in the coherent case the photon interacts with the full nucleus, while in the incoherent case it scatters off a single nucleon in the nucleus. In both cases a $\jpsi$ vector meson is produced.

The center-of-mass  energy of the photon-nucleus system is denoted by $\WgA$.  The square of the  four-momentum transfer in the nucleus vertex is denoted by $t$ and it is related to the transverse momentum ($p_T$) of the produced $\jpsi$, in the laboratory frame, by $-t =p^2_T$.
 The $t$ distribution is related to the transverse distribution of matter in the target. For the case of a large nucleus the $\jpsi$ produced in a coherent process has a transverse momentum of just a few tens of MeV/$c$, while in the incoherent process it reaches a few hundred of MeV/$c$, reflecting the difference in size between the nucleon and the nucleus.

\subsection{The nucleus-nucleus cross section}
The cross section $d\sigma_{\rm AA}/dy$ for the  photoproduction of a $\jpsi$ at rapidity $y$ in collisions of nuclei  can be factorized as the product of the photon flux $n_\gamma(y,\{\vec{b}\})$ and the  photonuclear  cross section $\sigma_{\gA}(y)$ as:

\begin{equation}
\frac{d\sigma_{\rm AA}}{dy} =
n_{\gamma}(y;\{\vec{b}\})\sigma_{\gA}(y)+
n_{\gamma}(-y;\{\vec{b}\})\sigma_{\gA}(-y),
\label{eq:XS}
\end{equation}
where  $\{\vec{b}\}$ delimits the range in impact-parameter $\vec{b}$ taken into account in the interaction. The two terms in this equation reflect the fact that any of the two nuclei can act as the source of the photon.

In the following, we will consider the rapidity $y$ in Eq. (\ref{eq:XS}) as given in the laboratory frame. It is defined with respect to the direction of the target.  The rapidity of the $\jpsi$ is related to the center-of-mass energy of the photon-nucleus system through
\begin{equation}
\WgA^2 = \sqrt{s_{\rm NN}}M_{\jpsi}e^{-y},
\end{equation}
where $M_{\jpsi}$ is the mass of the $\jpsi$ vector meson and $\sqrt{s_{\rm NN}}$  is the center-of-mass energy per nucleon pair in the A-A system.

The formalism for the computation of the flux has been described in detail in~\cite{Contreras:2016pkc}. The case for ultra-peripheral collisions (UPC) was originally proposed in~\cite{Klein:1999qj}. It is well known and it has been used extensively.
For the case of peripheral collisions and for the centrality class we are interested in (70--90\% centrality class) the formalism presented in~\cite{Contreras:2016pkc} produces numerically similar results as other proposals to compute the flux in peripheral collisions  as discussed in~\cite{Klusek-Gawenda:2015hja,Ducati:2017sdr}.

\subsection{The photon-nucleus cross section}
The photon-nucleus cross section appearing in Eq.~(\ref{eq:XS}) is given by the integral over $t$ of the following cross sections:
\begin{equation}
\left.
\frac{d\sigma_{\gA}}{dt}
\right|^{\rm coh}_{T,L} =
 \frac{(R^{T,L}_g)^2}{16\pi}
 \left|
\left< A^j(x,Q^2,\vec{\Delta})_{T,L}\right>_j
\right|^2
\label{eq:ExclXS}
\end{equation}
for the coherent process, and
\begin{eqnarray}
\left.
\frac{d\sigma_{\gA}}{dt}
\right|^{\rm inc}_{T,L} &=& \frac{(R^{T,L}_g)^2}{16\pi}
\left(\left<\left|
A^j(x,Q^2,\vec{\Delta})_{T,L}
\right|^2\right>_j  \right. \nonumber \\
&&-
\left.
\left|
\left< A^j(x,Q^2,\vec{\Delta})_{T,L}
\right>_j\right|^2\right)
\label{eq:DissXS}
\end{eqnarray}
for incoherent production, where the average is over different geometrical configurations $j$ of the nucleons,  respectively hot spots, inside the nucleus, as detailed below.

The amplitude is given by

\begin{eqnarray}
A^j(x,Q^2,\vec{\Delta})_{T,L} &=& i\int d\vec{r}\int^1_0\frac{dz}{4\pi}
(\Psi^*\Psi_{\rm V})_{T,L} \nonumber \\
& &\int d\vec{b}\;
e^{-i(\vec{b}-(1-z)\vec{r})\cdot\vec{\Delta}}
\left(\frac{d\sigma_{\rm dA}}{d\vec{b}}\right)_j,
\label{eq:Amplitude}
\end{eqnarray}
where  $\Psi_{\rm V}$ is the wave function of the vector meson, $\Psi$ is the wave function of a virtual photon fluctuating into a quark-antiquark color dipole, $\vec{r}$ is the transverse distance between the quark and the antiquark, and $z$ is the fraction of the longitudinal momentum of the dipole carried by the quark.

 The amplitude depends on the virtuality of the quasi-real photon $Q^2$, which we take to be $Q^2=0.05$ GeV$^2$. It also depends on $\vec{\Delta}^2=-t$ and on $x=(M_{\jpsi}/\WgA)^2$. The indices $T$ and $L$ represent the contributions of transversely and longitudinally polarized photons, respectively. The total cross section is the sum of the  $T$ and $L$ contributions.
 The QCD part of the model is contained in the dipole-target cross section $d\sigma_{\rm dA}/d\vec{b}$, which is discussed below.

The so-called skewedness correction~\cite{Shuvaev:1999ce} is given by
\begin{equation}
R^{T,L}_g(\lambda^{T,L}_g) = \frac{2^{2\lambda^{T,L}_g+3}}{\sqrt{\pi}}\frac{\Gamma(\lambda^{T,L}_g+5/2)}{\Gamma(\lambda^{T,L}_g+4)},
\label{eq:Rg}
\end{equation}
with $\lambda^{T,L}_g$ defined as
\begin{equation}
\lambda^{T,L}_g\equiv \frac{\partial\ln(A_{T,L})}{\partial\ln(1/x)}.
\label{eq:lambda}
\end{equation}

As we did in~\cite{Cepila:2016uku}, for $\Psi_{T,L}$ we use the definitions and parameter values of~\cite{GolecBiernat:1998js} and for the wave function of the vector meson, $\Psi_{\rm V}$,  we use the
boosted-Gaussian model ~\cite{Nemchik:1994fp,Nemchik:1996cw}, with the numerical values of the parameters  as in~\cite{Kowalski:2006hc}.

\subsection{The dipole-nucleus cross section}

We have followed two approaches to use the dipole-nucleon cross section determined in~\cite{Cepila:2016uku} to define the dipole-nucleus cross section without introducing any new additional free parameters.

In one approach, denoted GG in the figures, we chose the Glauber-Gribov methodology proposed in~\cite{Armesto:2002ny}, which relates the dipole-proton (dp) to the dipole-nucleus cross section via the nuclear profile $T_{\rm A}(\vec{b})$:

\begin{equation}
\left(\frac{d\sigma_{\rm dA}}{d\vec{b}} \right)_j= 2\left[
1-\exp\left(-\frac{1}{2}
\sigma_{\rm dp}(x,r)T^j_{\rm A}(\vec{b})\right)\right],
\end{equation}
where we use the  Golec-Biernat and Wusthoff~\cite{GolecBiernat:1998js} model for the dipole--proton cross section:
\begin{equation}
\sigma_{\rm dp}(x,r)= \sigma_0\left[1-\exp\left(-r^2Q^2_s(x)/4\right)\right],
\end{equation}
with $\sigma_0=\pi R^2_{\rm p}$, $R_{\rm p}$ the proton radius, and the saturation scale given by
\begin{equation}
Q^2_s(x) = Q^2_0(x_0/x)^\lambda,
\end{equation}
where the values of the  parameters are as in~\cite{Cepila:2016uku}.

In the second approach we use a model that factorizes  the $\vec{b}$ and $x$ dependences as proposed in~\cite{Cepila:2016uku}:
\begin{equation}
\left(\frac{d\sigma_{\rm dA}}{d\vec{b}} \right)_j=
\sigma^{\rm A}_0\left[1-\exp\left(-r^2Q^2_{A,s}(x)/4\right)\right]
T^j_{\rm A}(\vec{b}).
\end{equation}

Here, as in~\cite{Cepila:2016uku}, $\sigma^{\rm A}_0$ is related to the area of the target by $\sigma^{\rm A}_0=\pi R^2_{\rm A}$, where $R_{\rm A}$ is the ratio of the nucleus as defined in the corresponding Woods-Saxon distribution. The saturation scale of the nucleon is related, through geometric scaling, to the saturation scale of the nucleus as proposed in~\cite{Armesto:2004ud}:
\begin{equation}
Q^2_{s,{\rm A}}(x) = Q^2_s(x) \left(\frac{{\rm A}\pi R^2_{\rm p}}{\pi R^2_{\rm A}}\right)^{\frac{1}{\delta}},
\end{equation}
with $\delta=0.8$ as found in~\cite{Armesto:2004ud} and $\sigma_0$ fixed from the analysis in~\cite{Cepila:2016uku}. This model is denoted GS in the figures.

We explore two models for the nuclear profile, one considers the nucleus to be made up of nucleons and the second includes subnucleonic degrees of freedom in the form of hot spots. (In the figures these approaches are denoted by n and hs, respectively.) The key distinguishing feature of our model, with respect to that presented in~\cite{Mantysaari:2017dwh}, is that the number of hot spots in a given nucleon increases with decreasing $x$ as proposed in~\cite{Cepila:2016uku}.

In detail, we have for the first case that each nucleon has a Gaussian profile of width $B_{\rm p}$, centered at random positions $\{\vec{b}_i\}^j$ sampled from a Woods-Saxon nuclear profile for each configuration $j$:

\begin{equation}\label{eq:TAnucl}
T^j_{\rm A}(\vec{b}) = \frac{1}{2\pi B_{\rm p}}\sum^{\rm A}_{i=1}\exp\left(-\frac{(\vec{b}-\vec{b}^j_i)^2}{2B_{\rm p}}\right).
\end{equation}

For the hot-spot model we have that the nucleons are themselves made up of hot spots, distributed according to a Gaussian of width $B_{\rm hs}$.
\begin{equation}\label{eq:TAnuclHS}
T^j_{\rm A}(\vec{b}) = \frac{1}{2\pi B_{\rm hs}}\sum^{\rm A}_{i=1}\frac{1}{N_{\rm hs}}\sum^{N_{\rm hs}}_{k=1}
\exp\left(-\frac{(\vec{b}-\vec{b}^j_i-\vec{b}^j_k)^2}{2B_{\rm hs}}\right),
\end{equation}
where in this case $N_{\rm hs}$ is a random number drawn from a zero-truncated Poisson distribution, where the Poisson distribution has a mean value
\begin{equation}
\langle N_{\rm hs}(x) \rangle = p_0x^{p_1}(1+p_2\sqrt{x}).
\end{equation}

All the parameters of the model have been fixed from a comparison to data on $\jpsi$ photoproduction off protons~\cite{Cepila:2016uku}. The values of the parameters and the associated discussion can be found in~\cite{Cepila:2016uku} and will not be repeated here.

In the GS model, the integrals over impact parameter $\vec b$ are done analytically and
factorize from the rest of the integrals. This allowed us to use 10000 different profile
configurations, which increased the numerical precision of the computation to the
percent level.

On the other hand, the GG model has to be integrated numerically also over impact
parameter. We have used globally adaptive subdivision with importance sampling as
implemented in the Suave method \cite{Hahn:2004fe}. The program has been configured to reach
either 2\% precision or to stop after a given number of maximum tries. This last criteria
was needed because the convergence was slow. It has been checked that in no case the
numerical precision of the method exceeded 4\%. The previous discussion corresponds
to the numerical integration of one configuration. But the cross sections are proportional to the average
 (the variance) over configurations for the coherent (incoherent) process.
Due to the amount of computer resources needed in the GG model, we have only used
200 configurations for this case. This translates into a $\approx 2\%$ ($\approx 13 \%$) uncertainty for
the coherent (incoherent) results shown in Figs.~\ref{fig:XS_gPb_coh} and~\ref{fig:XS_gPb_inc}. Similar observations
for the fluctuations in the variance have already been noted in \cite{Toll:2012mb}. There is another
source of numerical uncertainty related to the fact that, to speed up the computation,
we used a grid in impact parameter space to store the profiles and then interpolations
have been used to obtain the specific values of the profile at the particular value of $\vec b$
needed at a given point. The effect of this approximation can be seen in Fig.~\ref{fig:XS_gPb_coh} as a
difference between the computation for the hot-spot and the nucleon profiles for the
coherent cross section.

\section{Predictions of the model and comparison to data}
\subsection{Results for the photonuclear cross sections}
\begin{figure}
\includegraphics[width=0.48\textwidth]{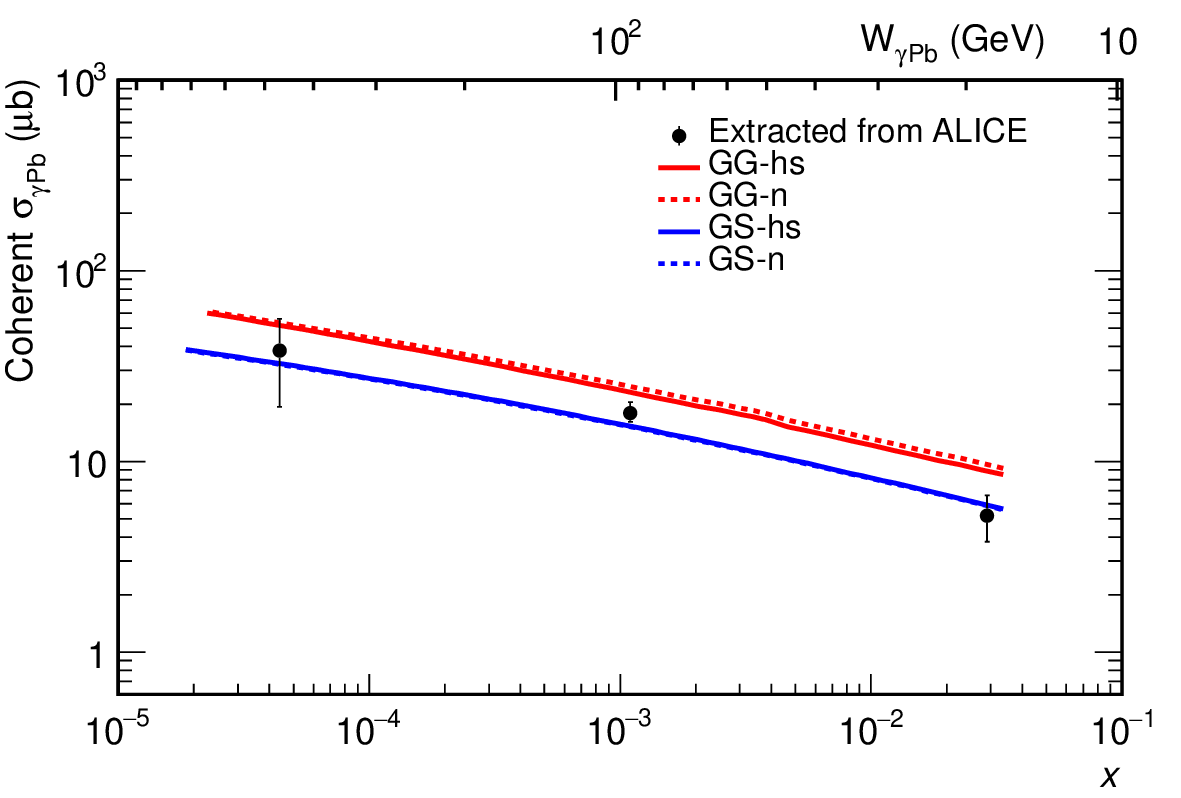}%
\caption{\label{fig:XS_gPb_coh}  Predictions for the coherent photonuclear production of $\jpsi$ off lead targets as a function of $x$ ($\WgPb$) compared with data extracted from ALICE measurements~\cite{Abelev:2012ba,Abbas:2013oua,Adam:2015gba}, as explained in~\cite{Contreras:2016pkc}.
}
\end{figure}
\begin{figure}
\includegraphics[width=0.48\textwidth]{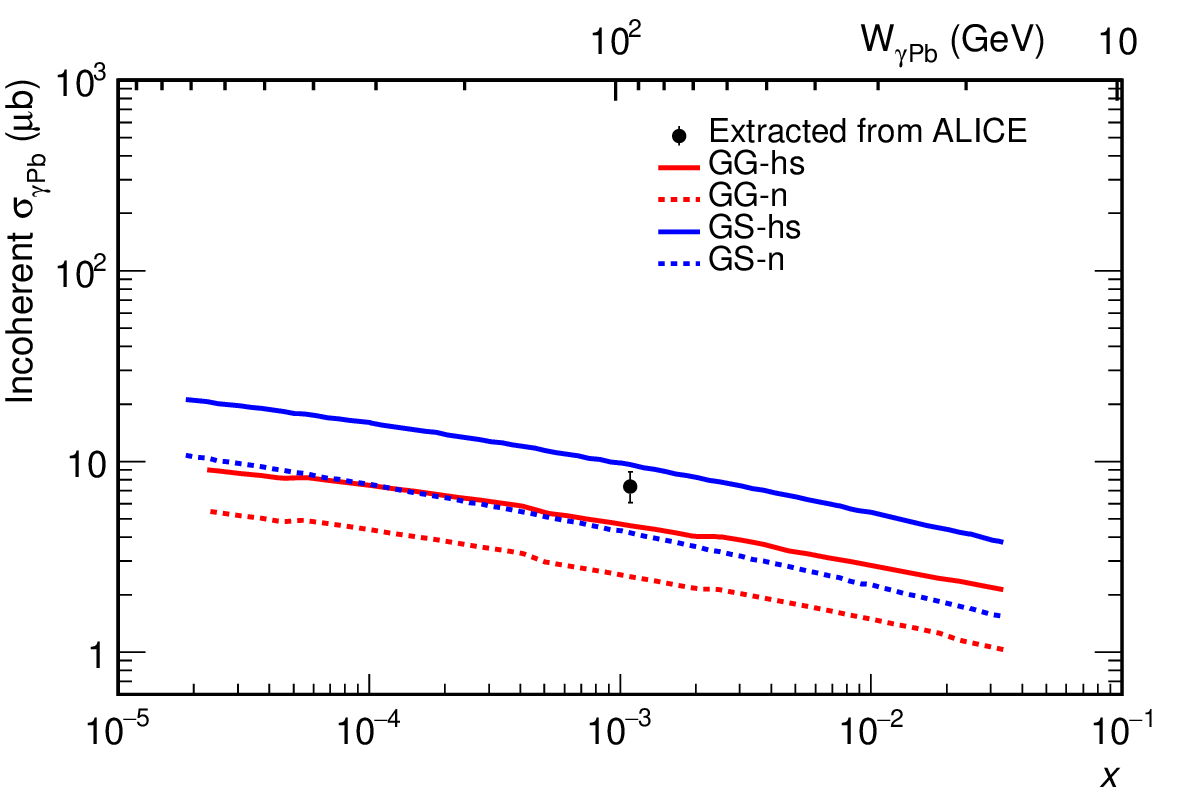}%
\caption{\label{fig:XS_gPb_inc}  Predictions for the incoherent photonuclear production of $\jpsi$ off lead targets as a function of $x$ ($\WgPb$) compared with data extracted from ALICE measurements~\cite{Abbas:2013oua}, as explained in~\cite{Contreras:2016pkc}.}
\end{figure}
\begin{figure}
\includegraphics[width=0.48\textwidth]{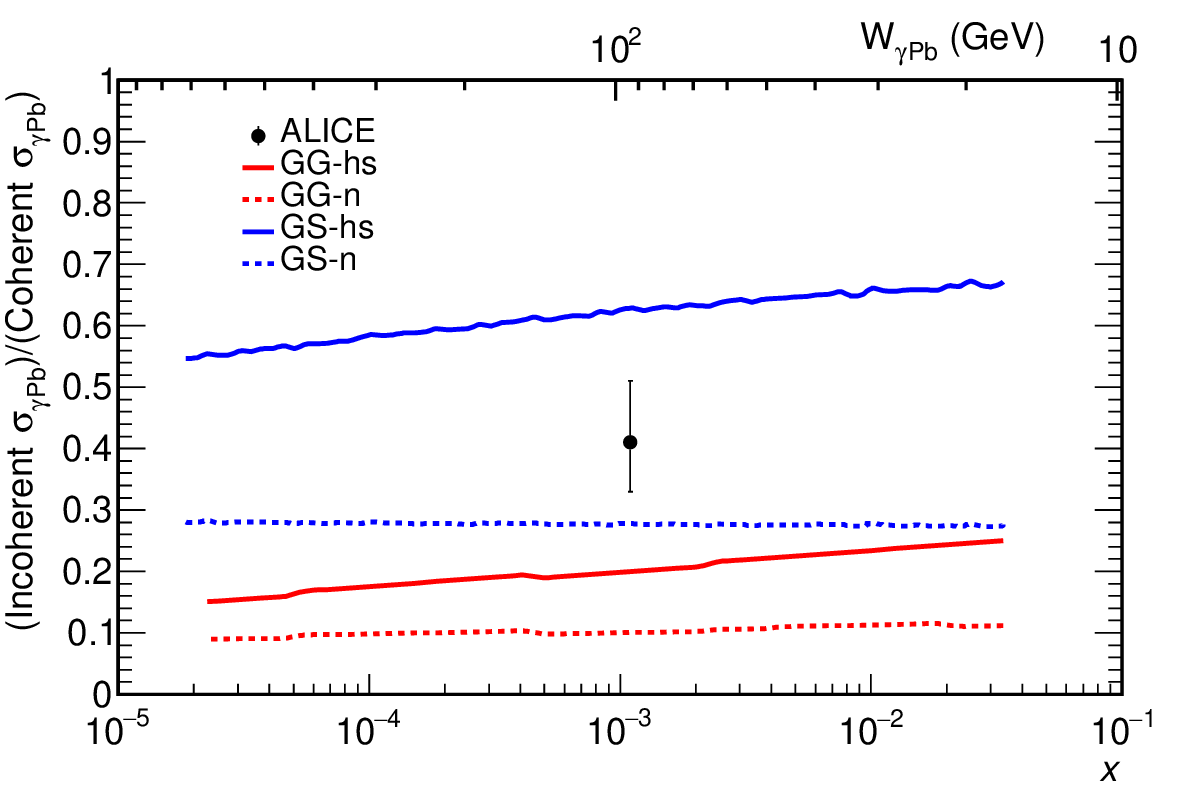}%
\caption{\label{fig:ratio_gPb}  Ratio of the incoherent to the coherent cross section for the photonuclear production of $\jpsi$ off lead targets as a function of $x$ ($\WgPb$) compared with  ALICE~\cite{Abbas:2013oua}. }
\end{figure}
\begin{figure}
\includegraphics[width=0.48\textwidth]{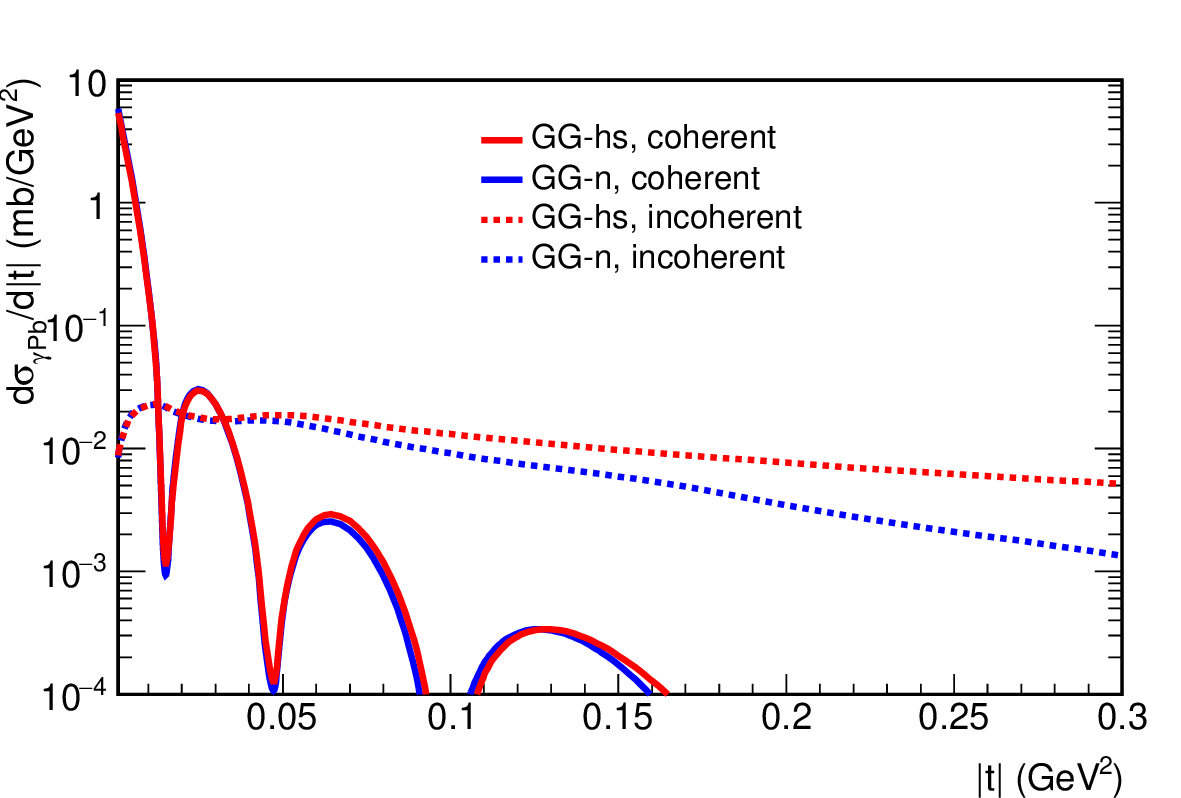}%
\caption{\label{fig:sigma_gPb_t}  Predictions for the $t$ dependence at mid-rapidity for the LHC Run 1 of the coherent and incoherent photonuclear cross section using the GG approach with (hs) or without (n) subnucleonic degrees of freedom. }
\end{figure}

\begin{table}
\caption{\label{tab:AuAu} Predictions of our model for the photonuclear production of $\jpsi$ and comparison with data from Au-Au collisions measured by PHENIX at $x=0.015$~\cite{Afanasiev:2009hy}. }
\begin{ruledtabular}
\begin{tabular}{lc}
Source & $\sigma_{\gamma\rm Au}$ ($\mu$b)  \\
\hline
PHENIX coherent~\cite{Afanasiev:2009hy} & 5.7 $\pm$ 2.3 (stat) $\pm$ 1.2 (syst)\\
PHENIX incoherent~\cite{Afanasiev:2009hy} & 3.6 $\pm$ 1.4 (stat) $\pm$ 0.7 (syst) \\
\hline
GS-hs coherent & 6.9 \\
GS-hs incoherent & 4.5 \\
GS-n coherent & 6.9 \\
GS-n incoherent & 1.8 \\
\hline
GG-hs coherent &  10.4 \\
GG-hs incoherent & 2.4 \\
GG-n  coherent & 11.2 \\
GG-n incoherent & 1.2 \\
\end{tabular}
\end{ruledtabular}
\end{table}

The predictions of our model for coherent photonuclear production of $\jpsi$ off lead are shown in Fig.~\ref{fig:XS_gPb_coh}. We present results for the case where the dipole-nucleus cross section was computed with the Glauber-Gribov approach  using either a nuclear profile made up of nucleons or of hot spots  (GG-n and GG-hs, respectively), as well as for the equivalent calculation performed within the geometric-scaling methodology (denoted by GS-n and GS-hs, respectively). These results are compared to data extracted from ALICE measurements in peripheral ~\cite{Adam:2015gba} and ultra-peripheral collisions~\cite{Abelev:2012ba,Abbas:2013oua} by taking into account the photon flux as explained in~\cite{Contreras:2016pkc}.

The predictions from the GS-hs and GS-n cases are very similar for coherent production and cannot be distinguished in the figure,
while in the GG approach there is a small difference in the prediction when considering
a nuclear structure made of hot spots or made of nucleons. The difference can be traced
back to the limited number of congurations explored and to the granularity of the
grid used to store the different profiles shown in Eqs.~(\ref{eq:TAnucl}) and~(\ref{eq:TAnuclHS}) as explained above.
The predictions of the GS approach give an excellent description of data, while the GG describe the lowest $x$ data, but  overestimate the measurements at larger values of $x$.

The predictions of our model for incoherent photonuclear production of $\jpsi$ off lead are shown in Fig.~\ref{fig:XS_gPb_inc}. In this case, there is a clear difference in the predictions when using a profile with subnucleonic degrees of freedom, with respect to the standard nuclear profile made of nucleons.  The comparison with the ALICE measurement~\cite{Abbas:2013oua} suggests that  data seems to prefer the hs models, specially for the GG computation. The same impression is obtained by looking at the ratio of the incoherent to coherent cross sections, as shown in Fig.~\ref{fig:ratio_gPb}. In addition, it can be seen in this figure, that the subnucleonic degrees of freedom introduce a dependence on $x$ for this ratio.

In order to exploit this $x$ dependence, we compare in Tab.~\ref{tab:AuAu} the results of our model for Au-Au collisions with data measured by PHENIX~\cite{Afanasiev:2009hy} at RHIC. This measurement was performed at mid-rapidity for a center-of-mass energy of the Au-Au system  $\sNN=0.2$ TeV, which corresponds to $x=0.015$. Even though the measurement has large uncertainties, PHENIX result, in conjunction with ALICE data, supports the prediction that the ratio of the incoherent to the coherent cross section for photonuclear production of $\jpsi$ depends on $x$. PHENIX data, as already seen in the case of ALICE, also seems to prefer the hot-spot models.

Figure~\ref{fig:sigma_gPb_t} shows the dependence of the coherent and incoherent cross sections on $t$ at mid-rapidity for the LHC Run 1 energies for the GG approach. A similar picture is obtained in the GS methodology. For the coherent case, the diffractive structure is clearly seen, while the incoherent cross section shows a large increase when including subnucleonic degrees of freedom in the nuclear profile, with respect to the case of a nucleon-only composition. Similar results have been shown in~\cite{Mantysaari:2017dwh}.

\subsection{Results for the Pb-Pb cross section}
\begin{figure}
\includegraphics[width=0.48\textwidth]{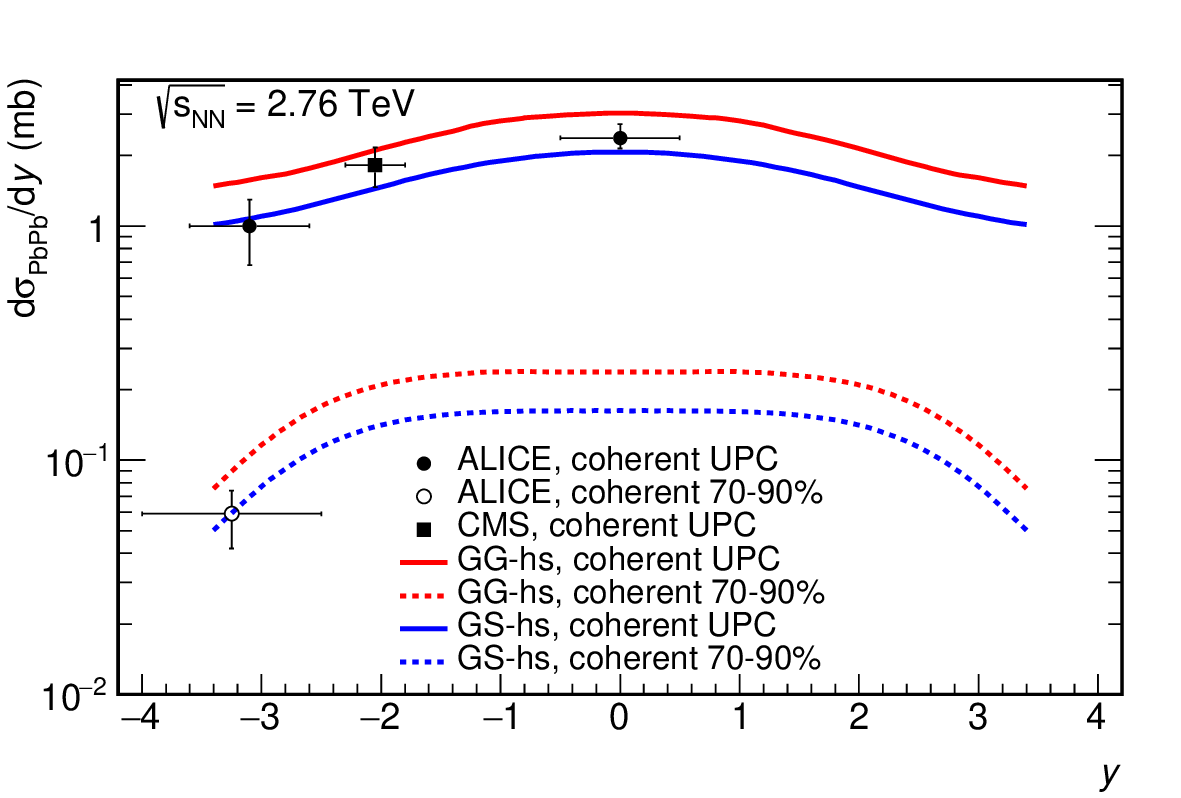}%
\caption{\label{fig:coh_AA}  Cross section for coherent photonuclear $\jpsi$ production in Pb-Pb collisions at $\sNN = 2.76$ TeV, corresponding to the LHC Run 1 energies, as a function of rapidity. Predictions of the model are compared to data measured in UPC by ALICE~\cite{Abelev:2012ba,Abbas:2013oua} and CMS~\cite{Khachatryan:2016qhq}, as well as data measured by ALICE~\cite{Adam:2015gba} in peripheral collisions in the 70-90\% centrality class.}
\end{figure}
\begin{figure}
\includegraphics[width=0.48\textwidth]{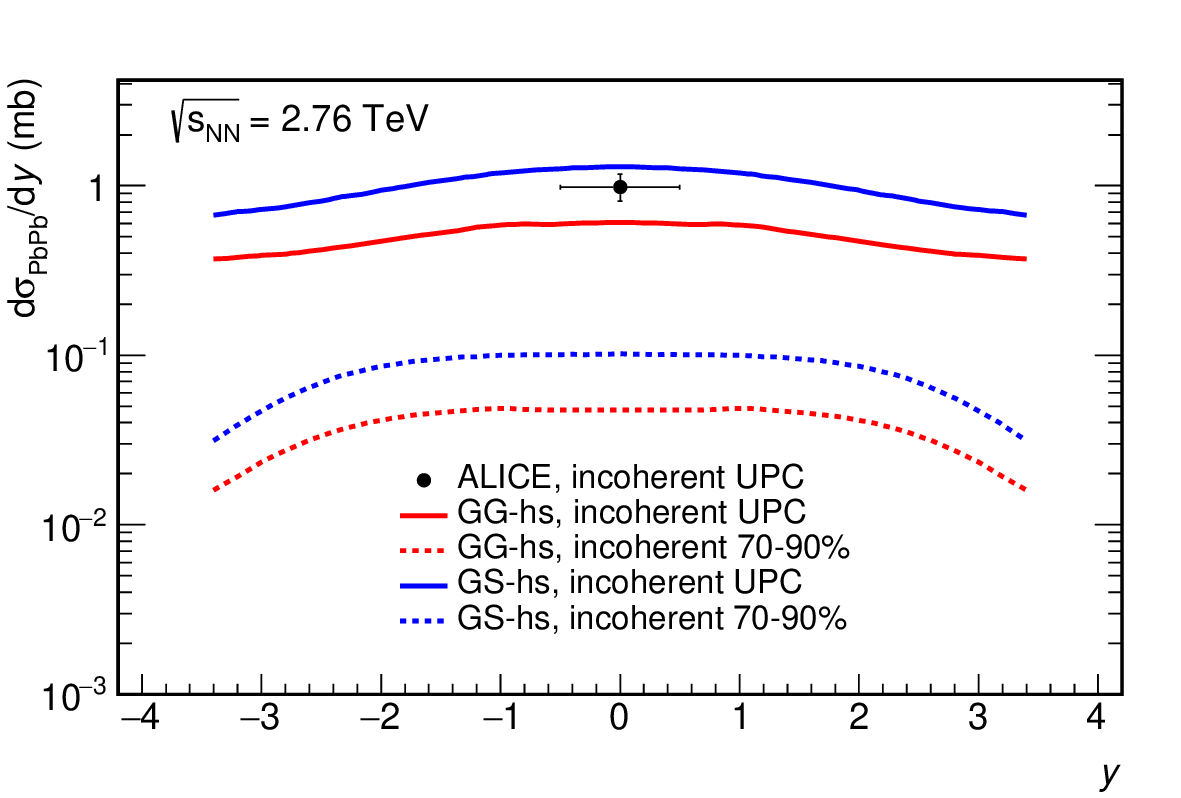}%
\caption{\label{fig:inc_AA}  Cross section for incoherent photonuclear $\jpsi$ production in Pb-Pb collisions at $\sNN = 2.76$ TeV, corresponding to the LHC Run 1 energies, as a function of rapidity. Predictions of the model are compared to data measured in UPC by ALICE~\cite{Abbas:2013oua}.}
\end{figure}
\begin{figure}
\includegraphics[width=0.48\textwidth]{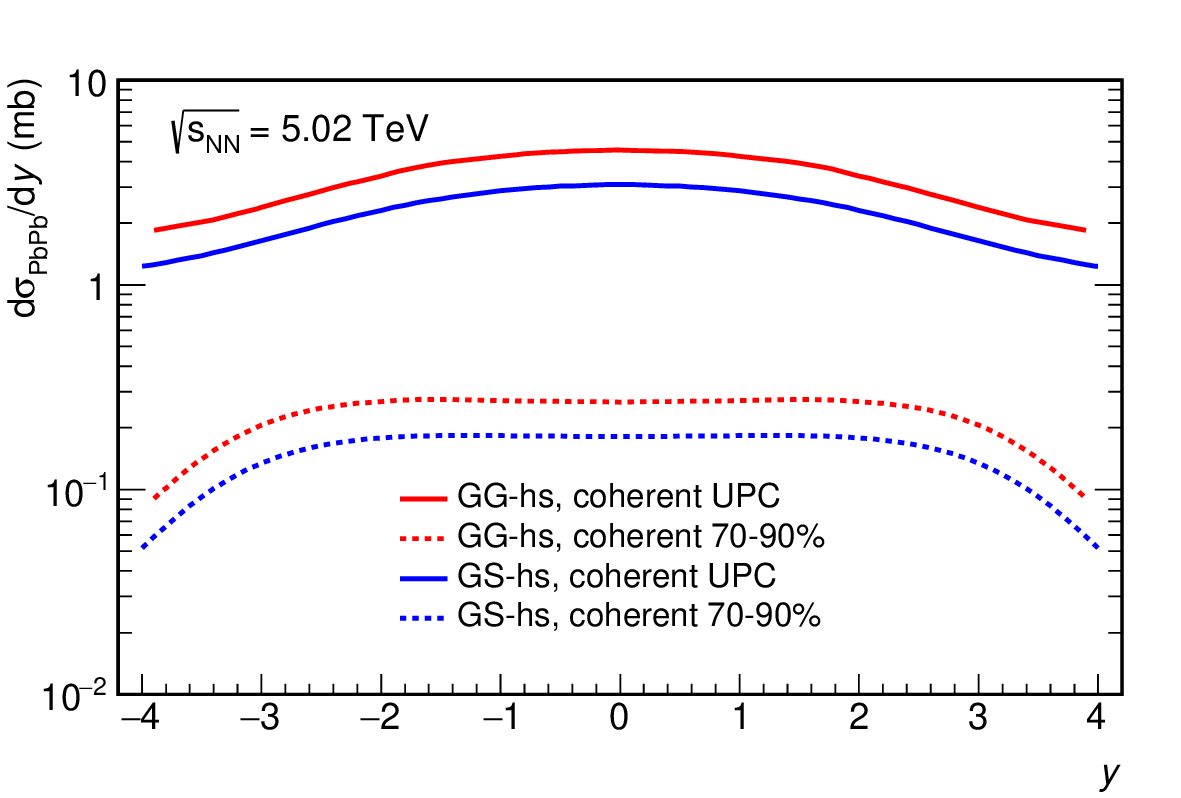}%
\caption{\label{fig:coh_AA2}  Cross section for coherent photonuclear $\jpsi$ production in Pb-Pb collisions at $\sNN = 5.02$ TeV, corresponding to the LHC Run 2 energies, as a function of rapidity. }
\end{figure}
\begin{figure}
\includegraphics[width=0.48\textwidth]{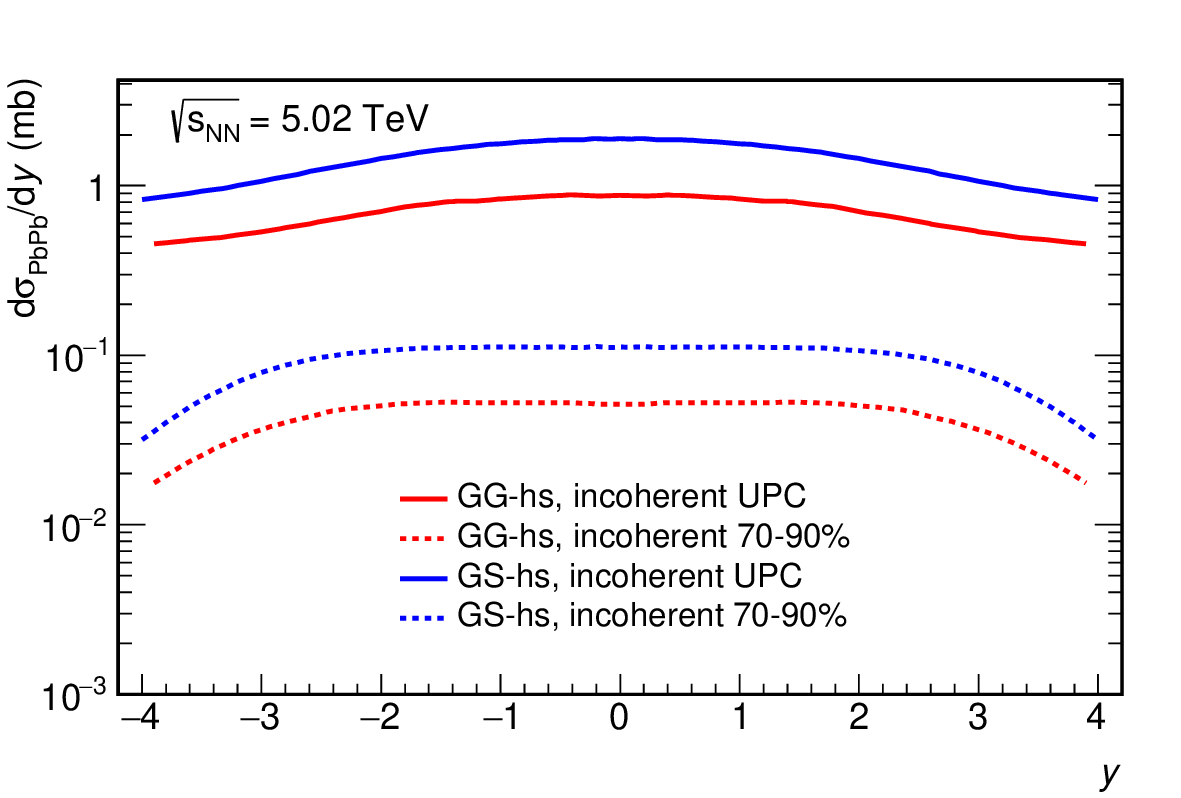}%
\caption{\label{fig:inc_AA2}  Cross section for coherent photonuclear $\jpsi$ production in Pb-Pb collisions at $\sNN = 5.02$ TeV, corresponding to the LHC Run 2 energies, as a function of rapidity. }
\end{figure}

The predictions of our model for  nucleus-nucleus cross sections, see Eq.~(\ref{eq:XS}), are shown in Figs.~\ref{fig:coh_AA} to~\ref{fig:inc_AA2}. The predictions for energies corresponding to the LHC Run 1 are compared to the available data. As of now there is no published data from the LHC Run 2 period for these observables.

In peripheral collisions,  we have used the results from ALICE~\cite{Abelev:2013qoq} to convert the centrality class to a range in impact parameter. According to which, the 70-90\% centrality class corresponds to a range (13.05,14.96) fm in impact parameter at Run 1 energies. The same formalism applied to LHC Run 2 energies yields  a range (13.1,15.0) fm.

Figure~\ref{fig:coh_AA}, shows that the GS-hs version of our model matches quite well the  data for coherent production available from ALICE~\cite{Abelev:2012ba,Abbas:2013oua} and CMS~\cite{Khachatryan:2016qhq} in UPC, as well as the data measured in peripheral interactions by ALICE~\cite{Abbas:2013oua}.
The GG-hs version of the model slightly overestimates the data.
A similar picture is obtained when comparing the prediction and the measurement for incoherent production as shown in Fig.~\ref{fig:inc_AA}.

Finally, Figs.~\ref{fig:coh_AA2} and~\ref{fig:inc_AA2} show the prediction of our model for Pb-Pb collisions at an energy corresponding to the LHC Run 2. All LHC collaborations already took Run 2 data in 2015 and a second data taking period is expected towards the end of 2018. These data, will produce measurements which are expected to have small experimental uncertainties.

 \section{Discussion}
The key new element of our computations, with respect to those presented in~\cite{Mantysaari:2017dwh} is the energy dependence of the hot-spot structure of the target in the impact-parameter plane. The parameters defining this dependence have been fixed in~\cite{Cepila:2016uku}. The most striking product of this difference is that the ratio of the incoherent to coherent cross section develops a dependence on energy, decreasing as the energy ($x$) is increase (decreased) as shown in Fig.~\ref{fig:ratio_gPb}.

Another difference of our model, with respect to that of~\cite{Mantysaari:2017dwh} is the implementation of the transition from proton to nuclear targets. The approach in~\cite{Mantysaari:2017dwh} is similar in spirit to the Glauber-Gribov model used here, but they use a different prescription to compute the dipole-proton interaction. Numerically the predictions from~\cite{Mantysaari:2017dwh} differ from ours. In particular, for the comparison with UPC data at mid rapidity, they overestimate the measurement for both, the coherent and the incoherent production, while in our case the GG-hs model overestimates the coherent process, but underestimates the incoherent production.

At larger rapidites, the model from~\cite{Mantysaari:2017dwh} agrees with the data from CMS in UPC slightly better than our predictions. They do not offer predictions at the largest rapidities measured at the LHC, because this region covers large values of $x$.

In phenomenological studies, it is customary to define the upper boundary of
the small $x$ region as $x \approx 0.01$. There is no strong quantitative argument to chose
exactly $x = 0.01$ as the upper limit for these studies. Here, we present results
including values of $x$ up to 0.034, which corresponds to a mild 25\% change in
$\ln(1/x)$ with respect to $x = 0.01$. This value was chosen to match the maximum
rapidity coverage of the ALICE detector for the LHC energies available in Run~2
 and to be able to compare to measurements from RHIC.

Another observation made in~\cite{Mantysaari:2017dwh} concerns the dependence of the results on the chosen
model for the wave function of the $\jpsi$, where the main effect between different models
is an overall change in normalization. In our model the normalization is given by the $\sigma_0$
parameter which has been fixed by construction to the effective area of the target. Once
the normalization was fixed, we checked in~\cite{Cepila:2016uku} that the same parameters described
correctly the behavior of inclusive deeply inelastic scattering - which do not depend on
the wave function of the vector meson - at the appropriate scale. This means that with
the current set of parameters in our model, a change of model for the wave function
of the $\jpsi$ would spoil the agreement with data, so in that sense all the components
of our model are fixed and we cannot change them. Note that it is expected that
the ratio of cross sections is less susceptible to these effects, so that it is a more solid
computation from the phenomenology side, as well as having smaller experimental
uncertainties.

The agreement between the predictions of our model and the experimental data is noteworthy, because it has been reached without adding any new free parameter to the model that successfully described the equivalent photoproduction processes off proton targets. In particular, the comparison of data with the predictions of the hot-spot models supports the existence of subnucleonic degrees of freedom in the nucleus and that this hot-spot structure evolves with $x$, respectively with $\WgA$.

Future measurements at RHIC and at the LHC are expected to have substantially lower experimental uncertainties providing valuable constraints to improve our picture of the nuclear structure. In particular, already existing data from UPC at mid rapidity from RHIC and at forward rapidities from the LHC-Run 2 energies, are dominated by contributions at large $x$. The measurement of the incoherent photonuclear production of $\jpsi$ in these  kinematic domains, will confirm or disprove the $x$ ($\WgA$) dependence of the incoherent to coherent cross section ratio.

\section{Summary, conclusions and outlook}
We have presented a model of the coherent and incoherent photoproduction of $\jpsi$ in hadronic targets, which includes an energy evolution of the QCD structure of the target in the impact-parameter plane. The parameters of the model had been fixed before by data from HERA and the LHC on proton targets. The model has been extended here to nuclear targets and compared to existing data from RHIC and LHC. The agreement between data and the model predictions is noteworthy, because it has been achieved without the addition of any new free parameters.

The main new ingredient of our approach, namely the inclusion of an energy evolution of the number of hot spots in the hadronic target, translates in this case in an energy dependence of the ratio of the incoherent to coherent cross sections. This prediction can be tested with already existing data from RHIC and the LHC Run 2. Future data from the LHC will provide new and stronger constraints to our model and help to understand better the subnucleonic structure of hadronic targets and its energy evolution in the small $x$ regime.

\section*{Acknowledgements}
We would like to thank Daniel Tapia-Takaki for useful discussions.
This work has been partially supported by the following grants:
Conicyt PIA/ACT 1406 (Chile), Conicyt PIA/Basal FB0821 (Chile)
and LTC17038 of the INTER-EXCELLENCE
program at the Ministry of Education, Youth and Sports (Czech Republic). \\

\bibliography{ALICE,QCD,HERA,CMS,ExpOthers}

\end{document}